\documentclass[twocolumn,aps,showpacs,preprintnumbers,amsmath,amssymb,superscriptaddress,floatfix,nofootinbib]{revtex4-1}
\usepackage{lipsum} 
\usepackage{graphicx} 
\usepackage{epsfig} 
\usepackage{epstopdf} 
\usepackage{hyperref}
\usepackage{amsmath} 
\usepackage{amsfonts} 
\usepackage{amssymb}
\usepackage[justification=justified,format=plain]{caption}
\usepackage{subcaption}
\DeclareMathSizes{12}{12}{12}{12}
\usepackage[section]{placeins}

\usepackage{float}

\usepackage[toc,page]{appendix}
\usepackage{slashed}
\usepackage{makecell}
\usepackage{feynman}
\usepackage[normalem]{ulem}
\usepackage{blindtext}
\usepackage{tikz}
\usetikzlibrary{arrows,shapes}
\usetikzlibrary{trees}
\usepackage{booktabs}
\usetikzlibrary{matrix,arrows} 				
\usetikzlibrary{positioning}				
\usetikzlibrary{decorations.pathmorphing}	
\usetikzlibrary{decorations.markings}
\tikzset{
vector/.style={decorate, decoration={snake}, draw},
provector/.style={decorate, decoration={snake,amplitude=2.5pt}, draw},
antivector/.style={decorate, decoration={snake,amplitude=-2.5pt}, draw},
fermion/.style={draw=black, postaction={decorate},
	decoration={markings,mark=at position .55 with {\arrow[draw=black]{>}}}},
fermionbar/.style={draw=black, postaction={decorate},
	decoration={markings,mark=at position .55 with {\arrow[draw=black]{<}}}},
fermionnoarrow/.style={draw=black},
gluon/.style={decorate, draw=black,
	decoration={coil,amplitude=4pt, segment length=5pt}},
scalar/.style={dashed,draw=black, postaction={decorate},
	decoration={markings,mark=at position .55 with {\arrow[draw=black]{>}}}},
scalarbar/.style={dashed,draw=black, postaction={decorate},
	decoration={markings,mark=at position .55 with {\arrow[draw=black]{<}}}},
scalarnoarrow/.style={dashed,draw=black},
electron/.style={draw=black, postaction={decorate},
	decoration={markings,mark=at position .55 with {\arrow[draw=black]{>}}}},
bigvector/.style={decorate, decoration={snake,amplitude=4pt}, draw},
}

\usepackage{tikz-feynman}

\begin{document}

\title{$\Xi_c$ and $\Xi_b$ excited states within a ${\rm SU(6)}_{\rm lsf}\times$HQSS model}

\author{J.~Nieves}
\affiliation{Instituto~de~F\'{\i}sica~Corpuscular~(centro~mixto~CSIC-UV),
  Institutos~de~Investigaci\'on~de~Paterna, Aptdo.~22085,~46071,~Valencia,
  Spain}
  
  \author{R.~Pavao}
\affiliation{Instituto~de~F\'{\i}sica~Corpuscular~(centro~mixto~CSIC-UV),
  Institutos~de~Investigaci\'on~de~Paterna, Aptdo.~22085,~46071,~Valencia,
  Spain}

  \author{L.~Tolos}
  \affiliation{Institut f\"ur Theoretische Physik, University of Frankfurt, Max-von-Laue-Str. 1, 60438 Frankfurt am Main, Germany}
  \affiliation{Frankfurt Institute for Advanced Studies, University of Frankfurt, Ruth-Moufang-Str. 1,
60438 Frankfurt am Main, Germany}
  \affiliation{Institute of Space Sciences (CSIC-IEEC), Campus Universitat Aut\`onoma de Barcelona, Carrer de Can Magrans, s/n, 08193 Cerdanyola del Vall\`es,
Spain} 

  \date{\today}

\begin{abstract}
We study odd parity $J=1/2$ and $J=3/2$ $\Xi_c$ resonances using a unitarized coupled-channel framework based on a ${\rm SU(6)}_{\rm lsf}\times$HQSS-extended Weinberg-Tomozawa baryon-meson interaction, while paying a special attention to the renormalization procedure. We predict a large molecular $\Lambda_c \bar K$ component for the $\Xi_c(2790)$ with a dominant $0^-$ light-degree-of-freedom spin configuration.  We discuss the differences between the $3/2^-$ $\Lambda_c(2625)$ and $\Xi_c(2815)$ states, and conclude that they cannot be SU(3) siblings, whereas we predict the existence of other  $\Xi_c-$states, two of them related to the two-pole structure of the $\Lambda_c(2595)$. It is of particular 
interest a pair of $J=1/2$ and $J=3/2$ poles, which form a HQSS doublet and that we tentatively assign to the $\Xi_c(2930)$ and $\Xi_c(2970)$, respectively. Within this picture, the $\Xi_c(2930)$ would be part of a SU(3) sextet, containing either the $\Omega_c(3090)$ or the $\Omega_c(3119)$, and that would be completed by the $\Sigma_c(2800)$. Moreover, we identify a $J=1/2$ sextet with the $\Xi_b(6227)$ state and  the recently discovered $\Sigma_b(6097)$. Assuming the {\it equal spacing rule} and to complete this multiplet, we predict the existence of a $J=1/2$ $\Omega_b$ odd parity state, with a mass of 6360 MeV and that should be seen in the $\Xi_b \bar K$ channel.
\end{abstract}
\maketitle


\section{Introduction}

The study of heavy baryons with charm or bottom content has been the subject of much interest over the past years in view of newly discovered states  \cite{Tanabashi:2018oca}. In particular, there has been a tremendous effort to understand the nature of the experimental states within conventional quarks models, QCD sum-rules frameworks, QCD lattice analysis or molecular baryon-meson models (see Refs.~\cite{Klempt:2009pi,Crede:2013sze,Cheng:2015iom,Chen:2016qju,Chen:2016spr,Guo:2017jvc} for recent reviews). 

The attention has been recently revived by the experimental observation of several excited states. Recent detections have been reported by the LHCb Collaboration regarding five $\Omega_c$ excited states in the $\Xi_c^+ K^-$ spectrum in $pp$ collisions \cite{Aaij:2017nav}, and the excited $\Xi_b(6227)$  state in $\Lambda_b^0 K^-$ and $\Xi_b^0 \pi^-$ invariant mass spectra also in $pp$ collisions \cite{Aaij:2018yqz}. Moreover, the Belle Collaboration has confirmed the observation of four of the excited $\Omega_c$ states  \cite{Yelton:2017qxg}, and detected the $\Xi_c(2930)$ state in its decay to $\Lambda_c^+ K^-$ in $B^- \rightarrow K^- \Lambda_c^+ \bar \Lambda_c^-$ decays \cite{Li:2017uvv}.

In view of these new observations, a large theoretical activity has been indeed triggered, in particular within dynamical approaches based on a molecular description of these states. 
Starting from the newly observed $\Omega_c$ states, the molecular models of Refs.~\cite{JimenezTejero:2009vq,Hofmann:2005sw} have been reanalyzed in view of the new discoveries. While in Ref.~\cite{Montana:2017kjw} two $\Omega_c$ resonant states at 3050 MeV and 3090 MeV with $J^P=1/2^-$ were obtained,  being identified with  two of the experimental states, in Ref.~\cite{Debastiani:2017ewu} two  $J^P=1/2^-$ $\Omega_c$ states  and one $J^P=3/2^-$ $\Omega_c$ were determined within an extended local hidden gauge approach, the first two in good agreement with \cite{Montana:2017kjw}.  Other theoretical works also examined the $\Omega_c$ sector, trying to explain the extra broad structure observed by the LHCb around 3188 MeV \cite{Aaij:2017nav}. In Ref.~\cite{Wang:2017smo} it was shown that this bump could be interpreted as the superposition of two $D \Xi$ bound states, whereas in  Ref.~\cite{Chen:2017xat} a loosely bound molecule of mass 3140 MeV was determined.

With regards to $\Xi_c$, the theoretical analysis based on the local hidden gauge formalism has shown that not only the $\Xi_c(2930)$ can have a molecular interpretation, but also other $\Xi_c$ states around 3 GeV reported in the PDG \cite{Tanabashi:2018oca}. In particular, the $\Xi_c(2790)$ would be a  $J^P=1/2^-$ molecular state, whereas $\Xi_c(2930)$, $\Xi_c(2970)$, $\Xi_c(3055)$ and $\Xi_c(3080)$ could be described as molecules with either $1/2^-$ or $3/2^-$ \cite{Yu:2018yxl}. On the other hand, the same model has produced two states for $\Xi_b(6227)$ with masses close to the experimental one with similar widths, being the spin-parity assignment either $1/2^-$ or $3/2^-$ \cite{Yu:2018yxl}. The $\Xi_b$ state has been also studied within a unitarized model that uses the leading-order chiral Lagrangian in Refs.~\cite{Lu:2014ina,Huang:2018bed}, identifying the $\Xi_b(6227)$ state as a $S$-wave $\Sigma_b \bar K$ molecule, with a preferred $1/2^-$ spin-parity assignment \cite{Huang:2018bed}.

Over the past years, a unitarized coupled-channel scheme has been developed in Refs.~\cite{GarciaRecio:2008dp,Gamermann:2010zz,Romanets:2012hm,GarciaRecio:2012db,Garcia-Recio:2013gaa,Tolos:2013gta,Garcia-Recio:2015jsa} that implements heavy-quark spin symmetry (HQSS), which is a proper QCD symmetry that appears when the quark masses, such as that of the  charm or bottom quark, become larger than the typical confinement scale. This approach is based on a consistent ${\rm SU(6)}_{\rm lsf} \times {\rm  HQSS}$ extension of the Weinberg-Tomozawa (WT) $\pi N$ interaction, where ``lsf'' stands for light quark-spin-flavor symmetry, respectively.  Within this framework, it has been identified a two-pole pattern for the $\Lambda_c(2595)$ resonance\footnote{The details of this double pole structure, generated by the $\Sigma_c\pi$, $ND$ and $ND^*$ coupled-channels dynamics, depend strongly on the adopted renormalization scheme, which could considerably enhance the role played by the two latter channels 
around the resonance energy. This is discuss in detail in Ref.~\cite{Nieves:2019nol}.} \cite{GarciaRecio:2008dp,Romanets:2012hm}, similar to the $\Lambda(1405)$ \cite{Oller:2000fj,Jido:2003cb,Hyodo:2007jq}. The same scheme  has also generated dynamically the $\Lambda_b(5912)$ and $\Lambda_b(5920)$ narrow resonances, discovered by LHCb \cite{Aaij:2012da}, which turn out to be HQSS partners, naturally explaining their approximate mass degeneracy~\cite{GarciaRecio:2012db}.

More recently, the work of Ref.~\cite{Romanets:2012hm} has been revisited in view of the newly discovered $\Omega_c$ states, paying a special attention to the renormalization procedure used in the unitarized coupled-channel model and its impact on the $\Omega_c$ sector. In Ref.~\cite{Nieves:2017jjx} it was shown that some (probably at least three) of the $\Omega_c$ states experimentally observed by LHCb would have $1/2^-$ or $3/2^-$. 

The discovery of the $\Xi_c(2930)$ and $\Xi_b(6227)$ has stimulated and motivated further research along this line. In the present work we follow a similar procedure as described in \cite{Nieves:2017jjx} and study the possible molecular interpretation of those states, revisiting the previous works on the $\Xi_c$ \cite{Romanets:2012hm} and $\Xi_b$ \cite{GarciaRecio:2012db} sectors. However, in the $\Xi_c$ sector we do not restrict ourselves to the recent $\Xi_c(2930)$ observation, but analyze all excited $\Xi_c$ states found experimentally with masses up to 3 GeV \cite{Tanabashi:2018oca}. The four excited $\Xi_c$ states with masses below 3 GeV and the $\Xi_b(6227)$ are collected in Table~\ref{tab:exp}, showing the assigned spin-parity $J^P$ (when possible) as well as  masses,  widths, and  decay channels. In this work we pay a special attention to the dependence on the renormalization scheme as well as  to the flavor-symmetry content of the ${\rm SU(6)}_{\rm lsf} \times {\rm  HQSS}$ model, as we determine the possible HQSS partners and siblings among the experimental states while predicting new ones. Thus, we follow the discussion of 
Ref.~\cite{Romanets:2012hm} on its spin-flavor symmetry breaking pattern. Flavor SU(4)
is not a good symmetry in the limit of a heavy charm quark,
for this reason, instead of the breaking pattern SU(8) $\supset$ SU(4), we consider the pattern
SU(8) $\supset$ SU(6), since the light spin-flavor group [SU(6)]
is decoupled from heavy-quark spin transformations. This
allows us to implement HQSS in the analysis and to unambiguously identify the corresponding
multiplets among the resonances generated dynamically.
At the same time, we are also able to assign
approximate heavy [SU(8)] and light [SU(6)] spin-flavor
multiplet labels to the states.

This work is organized as follows. In Section \ref{formalism} we present the ${\rm SU(6)}_{\rm lsf} \times {\rm HQSS}$
extension of the WT interaction, while in Section \ref{sec:res} we show our results for the $\Xi_c$  and $\Xi_b$ states\footnote{From now on we refer to excited $\Xi_c$ and $\Xi_b$ independently of $1/2^-$ or $3/2^-$ spin-parity assignment}, respectively, and the possible experimental identification. Finally, in Section \ref{conc} we present our conclusions, emphasizing the possible classification of these experimental states according to the flavor-symmetry content of the scheme, while predicting new observations.

\section{Formalism}
\label{formalism}

\begin{widetext}

\begin{table}[H]
\begin{tabular}{c|c|c|c|c}
${\rm {\bf Baryon}}$ & $J^P$ & ${\rm {\bf M \ (MeV)}}$ & {\bf $\Gamma$ (MeV)} & ${\rm {\bf Decay \ channels}}$  \\
$\Xi_c(2790)^+/\Xi_c(2790)^0$ & $1/2^-$ & $2792.4 \pm 0.5  \ / \ 2794.1 \pm 0.5 $ & $ 8.9 \pm 1.0 \ /  \ 10.0 \pm 1.1$& $\Xi_c' \pi$ \\ 
$\Xi_c(2815)^+/\Xi_c(2815)^0$ &  $3/2^-$ & $2816.73 \pm 0.21 \ /  \ 2820.26 \pm 0.27 $ & $2.43 \pm 0.26 \ / \ 2.54 \pm 0.25$ & $\Xi_c' \pi$, $\Xi_c^* \pi$ \\ 
 $\Xi_c(2930)^+/ \Xi_c(2930)^0$ & $?$ & $2942 \pm 5 \ / \ 2929.7 ^{+2.8}_{-5.0}$ &  $15 \pm 9 \ / \ 26 \pm 8$ & $\Lambda_c^+ K^-$, $\Lambda_c^+ K_S^0$ \\ 
  $\Xi_c(2970)^+ / \Xi_c(2970)^0$ & $?$ & $2969.4 \pm 0.8 \ / \ 2967.8^{+0.9}_{-0.7}$ & $20.9^{+2.4}_{-3.5} \ / \ 28.1 ^{+3.4}_{-4.0}$ & $\Lambda_c^+ \bar K \pi$, $\Sigma_c \bar K$, $\Xi_c 2 \pi$, $\Xi_c' \pi$, $\Xi_c^* \pi$ \\ 
  $\Xi_b(6227)$ & $?$ & $6226.9 \pm 2$ & $18 \pm 6$ & $\Lambda_b^0 K^-$, $\Xi_b^0 \pi^-$ 
\end{tabular}
\caption{Excited $\Xi_c$ states below 3 GeV and the excited $\Xi_b$ state found experimentally \cite{Tanabashi:2018oca}. We show the assigned $J^P$ (when possible), the mass ${\rm M}$ and width $\Gamma$, as well as the decay channels.} 
\label{tab:exp}
\end{table}
\end{widetext}

We consider the sector with charm $C=1$, strangeness $S=-1$ and isospin $I=1/2$ quantum numbers, where the
$\Xi_c(2930)$ state has been observed by the Belle Collaboration \cite{Li:2017uvv}. Also, we examine the bottom $B=-1$, strangeness $S=-1$ and isospin $I=1/2$, where the $\Xi_b(6227)$ has been found \cite{Aaij:2018yqz}. In order to do so, we revise the results in Ref.~\cite{Romanets:2012hm} for the $\Xi_c$ states and in Ref.~\cite{GarciaRecio:2012db} for $\Xi_b$ ones.

In the case of  the $C=1$, $S=-1$ and $I=1/2$ sector, the building-blocks are the pseudoscalar ($D_s, D, K, \pi,\eta, {\bar K}$) and vector ($D_s^*,
D^*,K^*, \rho,\omega, {\bar K}^{*}, \phi$) mesons, and the spin-1/2 ($\Lambda$, $\Sigma$, $\Xi$, $\Lambda_c$,
$\Sigma_c$, $\Xi_c$, $\Xi'_c$, $\Omega_c$), and spin-3/2 ($\Sigma^*_c$, $\Xi^*_c$, $\Omega_c^*$) charmed
baryons~\cite{Romanets:2012hm,GarciaRecio:2008dp}. For the bottom sector $B=-1$, $S=-1$ and $I=1/2$, one can substitute the $c$ quark by a $b$ quark, and we have the pseudoscalar ($\bar B_s, \bar B,
K, \pi,\eta, {\bar K}$) and vector ($\bar B_s^*,\bar B^*,K^*, \rho,\omega, {\bar K}^{*}, \phi$) mesons, and the spin-1/2 ($\Lambda$, $\Sigma$, $\Xi$, $\Lambda_b$, $\Sigma_b$, $\Xi_b$, $\Xi'_b$, $\Omega_b$), and spin-3/2
($\Sigma^*_b$, $\Xi^*_b$, $\Omega_b^*$) baryons~\cite{GarciaRecio:2012db}. All baryon-meson pairs with
$(C=1/B=-1, S=-1, I=1/2)$ quantum numbers span the coupled-channel space for a given total angular momentum ($J$).

The $S$-wave tree level amplitudes between two baryon-meson channels are given by the ${\rm SU(6)}_{\rm lsf} \times {\rm HQSS}$ WT  kernel,
\begin{equation}
\label{eq:WT}
V_{ij}^J(s) = D_{ij}^J \frac{2 \sqrt{s}-M_i-M_j}{4 f_i f_j} \sqrt{\frac{E_i+M_i}{2 M_i}} \sqrt{\frac{E_j+M_j}{2M_j}}.
\end{equation}
The $M_i$ and $m_i$ are the masses of the baryon and meson in the $i$
channel, respectively, and $E_i$ is the center-of-mass
energy of the baryon in the same channel,
\begin{equation}
E_i=\frac{s-m_i^2+M_i^2}{2 \sqrt{s}}.
\end{equation}
The hadron masses, meson decay constants, $f_i$, and $D_{ij}^J$ matrices are taken from
Ref.~\cite{Romanets:2012hm,GarciaRecio:2012db}, where the underlying ${\rm SU(6)}_{\rm lsf} \times$ HQSS group
structure of the interaction has been considered.

Starting from $V^J_{ij}$, we solve the Bethe-Salpeter equation (BSE) in coupled channels, 
\begin{equation}
\label{eq:LS}
T^J(s)=\frac{1}{1-V^J(s) G^J(s)} V^J(s),
\end{equation}
where the  $G^J(s)$ is a diagonal matrix that contains the different baryon-meson loop functions $G_i$,
\begin{equation}
\label{eq:normloop}
G_i(s)=i 2M_i  \int \frac{d^4 q}{(2 \pi)^4} \frac{1}{q^2-m_i^2+i\epsilon} \frac{1}{(P-q)^2-M_i^2+i\epsilon},
\end{equation}
with $P$ the total momentum of the system such that $P^2=s$. We omit
the index $J$ from here on for simplicity. The bare loop function is
logarithmically ultraviolet (UV) divergent and needs to be
renormalized. This can be done by separating the divergent and finite parts of the loop function,
\begin{equation}
G_i(s)=\overline{G}_i(s)+G_i(s_{i+}) ,
\label{eq:div}
\end{equation}
with the finite part of the loop function, $\overline{G}_i(s)$, given in Refs.~\cite{Nieves:2001wt} and \cite{Nieves:2017jjx}. The divergent contribution of the loop function, $G_i(s_{i+})$ in
Eq.~(\ref{eq:div}) needs to be renormalized. 

On the one hand, this can be done by one subtraction at certain scale ($\sqrt{s}=\mu$)
\begin{eqnarray}
G_i^\mu(s) =\overline{G}_i(s) - \overline{G}_i(\mu^2) ,
\label{eq:relation}
\end{eqnarray}
where $\mu = \sqrt{m_{th}^2+M_{th}^2}$, with $m_{th}$ and $M_{th}$ the masses of the meson and the baryon, respectively, that belong to the channel with the smallest threshold for a given $(C,S,I)$ or $(B,S,I)$ sectors.   This common scale $\mu$ is chosen to be  independent
of total angular momentum $J$ \cite{Hofmann:2005sw, Hofmann:2006qx}, and it is the approach used in the previous works of Refs.~\cite{Romanets:2012hm,GarciaRecio:2012db}.

On the other hand, as discussed in our recent paper \cite{Nieves:2017jjx}, we could also use a sharp-cutoff regulator $\Lambda$ in momentum space, so  that
\begin{equation}
G^{\Lambda}_i(s) =\overline{G}_i(s) + G_i^{\Lambda}(s_{i+}), \label{eq:uvcut2}
\end{equation}
where $G_i^{\Lambda}(s_{i+})$  is given in Refs.~\cite{GarciaRecio:2010ki, Nieves:2017jjx}.

Note that if one uses channel-dependent cutoffs, the one-subtraction renormalization scheme  is recovered by
choosing  $\Lambda_i$ in each channel in such a way that
\begin{equation}
G^{\Lambda_i}_i(s_{i+})= -\overline{G}_i(\mu^2) .
\label{eq:subtraction}
\end{equation}
However, we employ a common UV cutoff for all baryon-meson loops within reasonable limits. In this manner, we avoid a fictitious reduction or enhancement of any baryon-meson loop by using an unappropriated  value of the cutoff~\cite{Guo:2016nhb, Albaladejo:2016eps, Nieves:2019nol}, as well as we prevent an arbitrary variation of the subtraction constants, as we correlate all of them with a UV cutoff~\cite{Nieves:2017jjx}. 

The poles of the $T$ matrix describe the odd-parity dynamically-generated $\Xi_c$ and $\Xi_b$ states, which appear in the first and second Riemann sheets (FRS and SRS). Poles of the scattering amplitude on the FRS below threshold are bound states, whereas poles on the SRS below the real axis and above threshold are resonances. The mass and the
width of the bound state/resonance can be found from the position of the pole on the complex energy plane. Close to the pole, the $T$-matrix behaves
as 
\begin{equation}
\label{eq:T-mat}
T_{ij}(s) \simeq \frac{g_i g_j}{\sqrt{s}-\sqrt{s_R}},
\end{equation}
where $\sqrt{s_R}=M_R - \rm{i}\, \Gamma_R/2$, with $M_R$ the mass and $\Gamma_R$ the width of the state, and $g_i$ is the complex coupling of the state to the channel $i$.
The dimensionless couplings $g_i$ are obtained by first assigning an arbitrary sign to one of them ($g_1$), so
\begin{equation}
g_1^2=\lim_{\sqrt{s}\rightarrow\sqrt{s_R}} (\sqrt{s}-\sqrt{s_R})T_{11}(s).
\end{equation}
The other couplings are calculated as,
\begin{equation}
g_j = g_1 \lim_{\sqrt{s}\rightarrow\sqrt{s_R}} \frac{T_{1j}(s)}{T_{11}(s)} ,
\end{equation}
and can be used to analyze the contribution of each baryon-meson channel to
the generation of the state. 

\vspace{0.5cm}

\section{Results}
\label{sec:res}

\subsection{ $\Xi_c$ excited states}


\begin{widetext}

\begin{figure}[H]
\centering
\begin{subfigure}[b]{0.7\textwidth}
\centering
\includegraphics[width=1.2\textwidth]{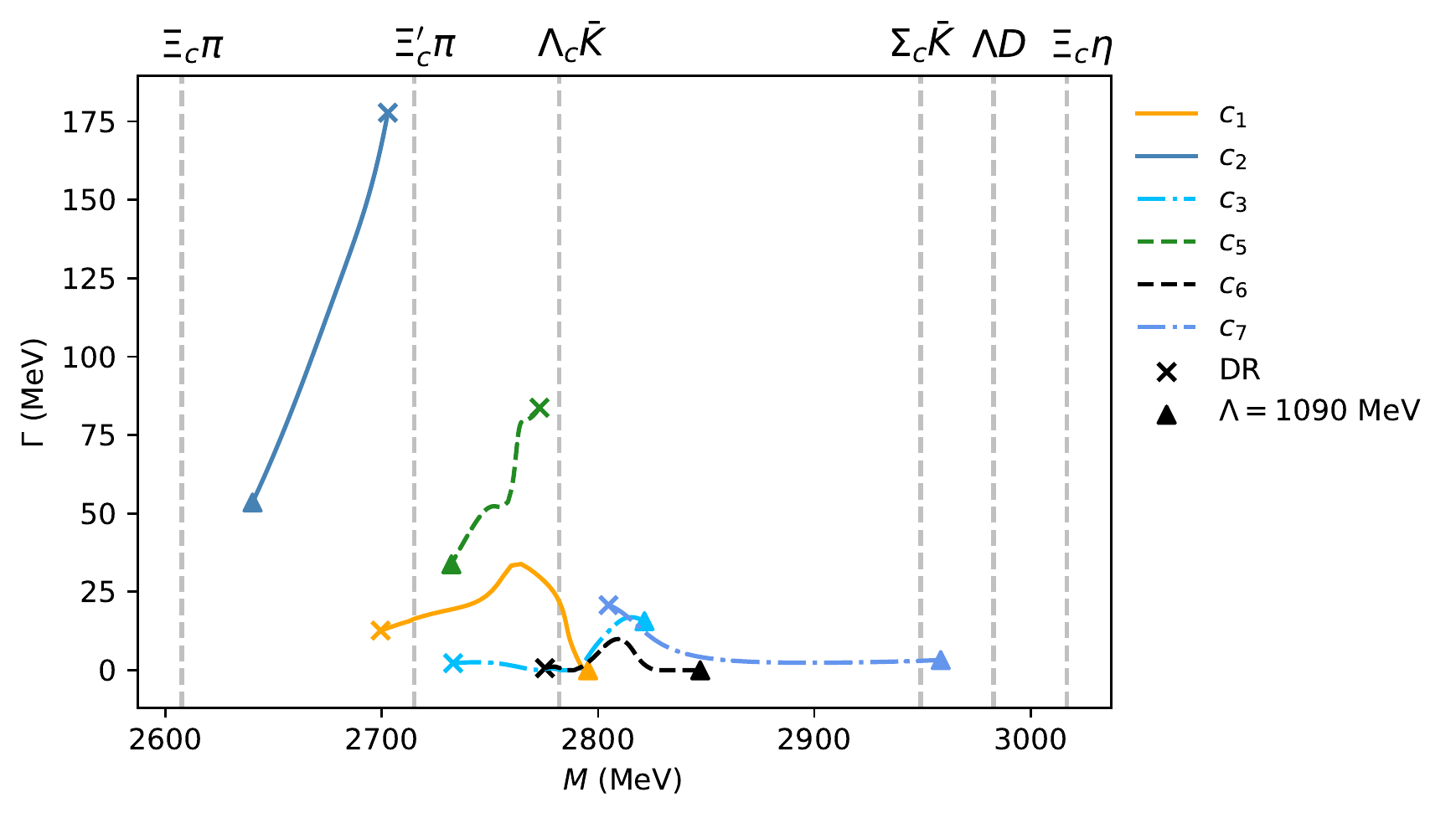}
\caption{$J=1/2$}
\label{fig:ctodim_12}
\end{subfigure}

\begin{subfigure}[b]{0.7\textwidth}
\centering
\includegraphics[width=1.2\textwidth]{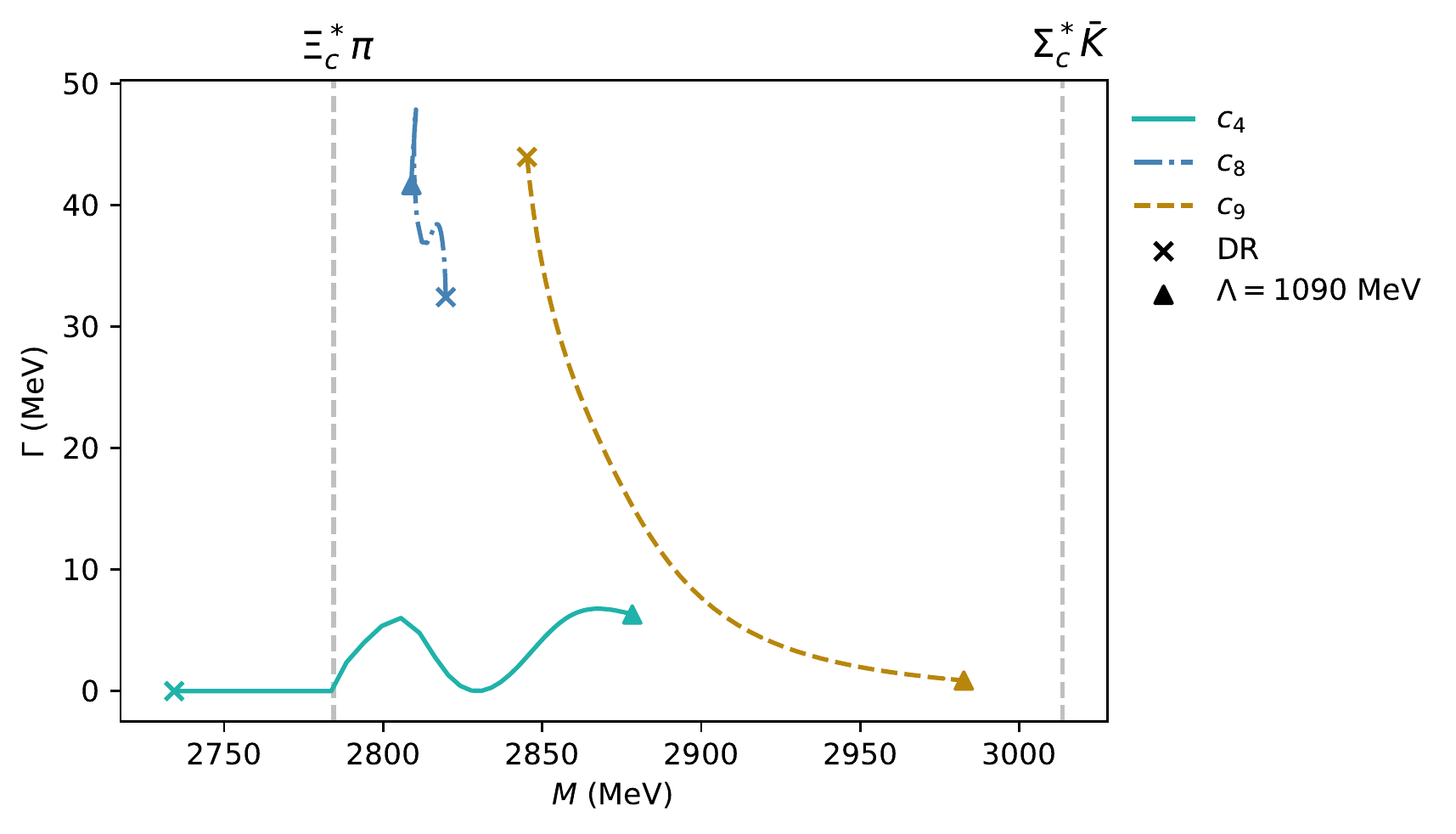}
\caption{$J=3/2$}
\label{fig:ctodim_32}
\end{subfigure}
\caption{Evolution of the masses and widths of the dynamically generated $\Xi_c$ states as we vary the renormalization scheme from using a subtraction constant to a common cutoff of $\Lambda=$1090 MeV. The cross symbolizes the position of the states in the subtraction constant scheme (or dimensional regularization (DR))~\cite{Romanets:2012hm}, while the triangle indicates the mass and width of the same states for the cutoff scheme. }
\label{fig:ctodim}
\end{figure}

\end{widetext}

As mentioned in the Introduction, the first observation of the $\Xi_c(2930)$ state was reported by the Belle Collaboration in Ref.~\cite{Li:2017uvv}. This state  was observed through its decay to  $\Lambda_c^+ K^-$ with no assigned quantum numbers.  Besides this recently discovered state, there are other three $\Xi_c$ excited states with energies below 3 GeV \cite{Tanabashi:2018oca}. As seen in Table~\ref{tab:exp}, the $1/2^-$ $\Xi_c(2790)$ state decays into $\Xi_c' \pi$, whereas the $3/2^-$ $\Xi_c(2815)$ decays into $\Xi_c' \pi$ and has also the decay chain $\Xi_c^* \pi$, followed by $\Xi_c^* \to \Xi_c \pi $~\cite{Yelton:2016fqw}.  Also, a $\Xi_c(2970)$ with unknown quantum numbers  has been observed decaying into $\Lambda_c^+ \bar K \pi$, $\Sigma_c \bar K$,  $\Xi_c 2 \pi$,  $\Xi_c' \pi$ and $\Xi_c^* \pi$.

We start by revising the results Ref.~\cite{Romanets:2012hm} in the $\Xi_c$ sector in order to understand whether the experimental states can be accommodated in our model. The widths of our $\Xi_c$ states as a function of their masses in the $J=1/2$ and $J=3/2$ sectors are shown in the upper and lower plots of Fig.~\ref{fig:ctodim}, respectively, together with different baryon-meson thresholds, to which they can couple. The dynamically generated states of Ref.~\cite{Romanets:2012hm}  are displayed with a cross and the "DR" legend, as those have been obtained using one subtraction at certain scale or dimensional regularization. In what follows, we label the states as $c_1\dots c_9$, and they correspond to those given in Table V of 
Ref.~\cite{Romanets:2012hm}. They have either $J^P=1/2^-$ or $J^P=3/2^-$ and are ordered by their mass position. Hence, $c_1$ ($c_9$) corresponds to the lightest (heaviest) state of mass 2699.4 MeV (2845.2 MeV), among those quoted in the mentioned table, where their SU(6) and SU(3) quantum numbers are also given.
We observe that the masses of our $\Xi_c$ states using one subtraction constant (DR) are below or close to the experimental $\Xi_c(2790)$ or $\Xi_c(2815)$ states, while being far below in mass with respect to $\Xi_c(2930)$ or $\Xi_c(2970)$.  

We then study the effect on masses and widths of the renormalization procedure so as to determine whether any our $\Xi_c$ can be identified with a experimental state while assessing the dependence on the renormalization scheme, which might be significant (as shown in Ref.~\cite{Nieves:2019nol}). We proceed as described in Ref.~\cite{Nieves:2017jjx} for the $\Omega_c$ states, where we explore a different renormalization procedure, the cutoff scheme. In order to do so, we first need to determine how the masses and widths of our dynamically generated states change as we adiabatically move from the one subtraction renormalization scheme to the cutoff one. Thus, we change the loop functions by 
\begin{equation}
\label{eq:adiab2}
G^{\Lambda}_i(s) =\overline{G}_i(s) - (1-x) \ \overline{G}_i(\mu^2) + x \ \overline{G}^{\Lambda}_i(s),
\end{equation}
where we slowly evolve $x$ from 0 to 1 while following the evolution of the states, as seen in Fig.~\ref{fig:ctodim}. The  $c_1$ to $c_9$ states for a $\Lambda=1090$ MeV are shown with a triangle. We find that most of these states move to higher energies, except for $c_2$, $c_5$ and $c_8$, whereas getting closer to the experimental values. Note that fot this cutoff, the $J^P = 1/2^-$ $c_1$ state become virtual above the $\Lambda_c \bar K$ threshold.

\begin{widetext}

\begin{figure}[H]
\centering
\begin{subfigure}[t]{0.7\textwidth}
\centering
\includegraphics[width=1.2\textwidth]{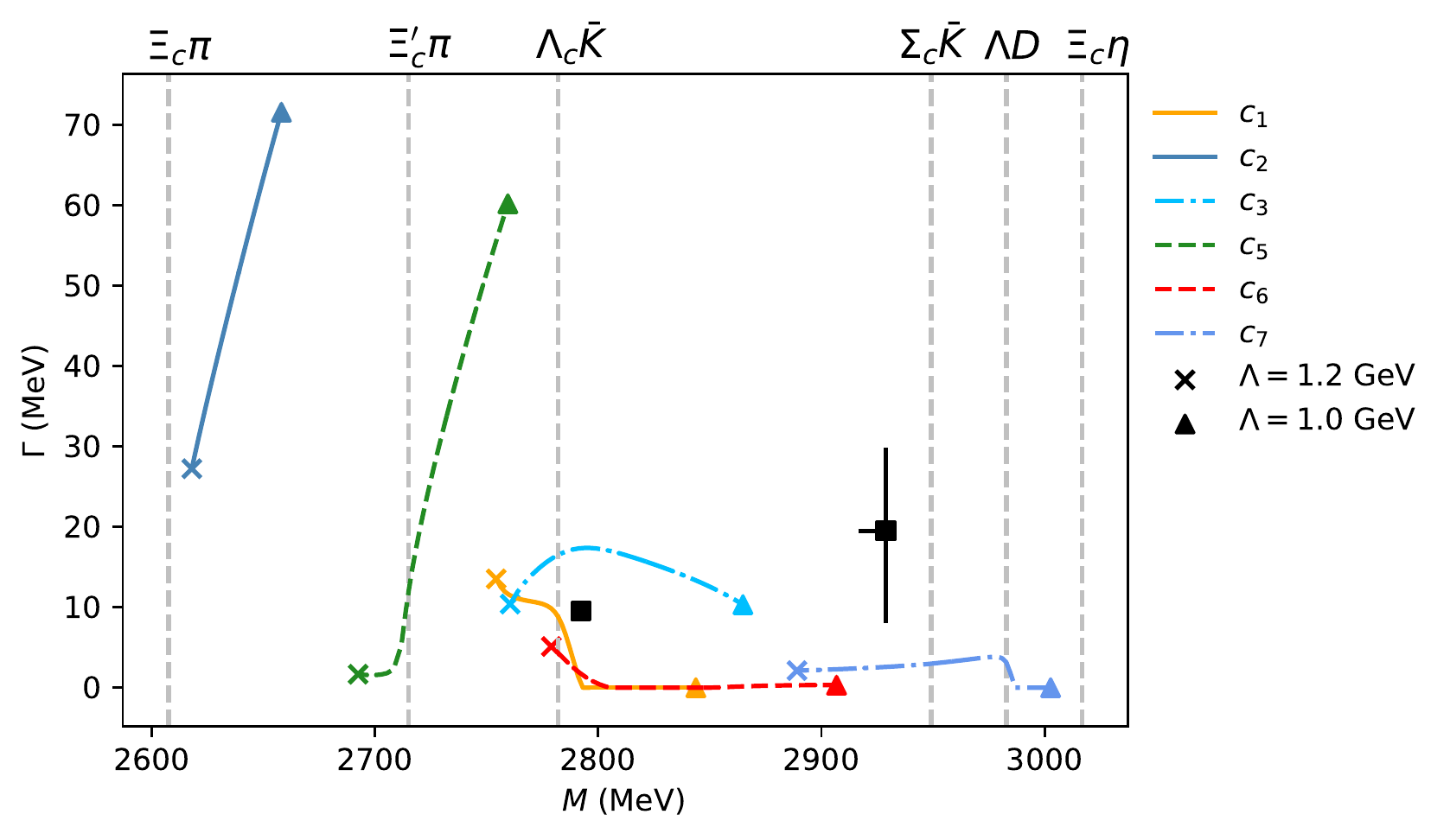}
\caption{$J=1/2$}
\label{fig:traj-c-12}
\end{subfigure}

\begin{subfigure}[t]{0.7\textwidth}
\centering
\includegraphics[width=1.2\textwidth]{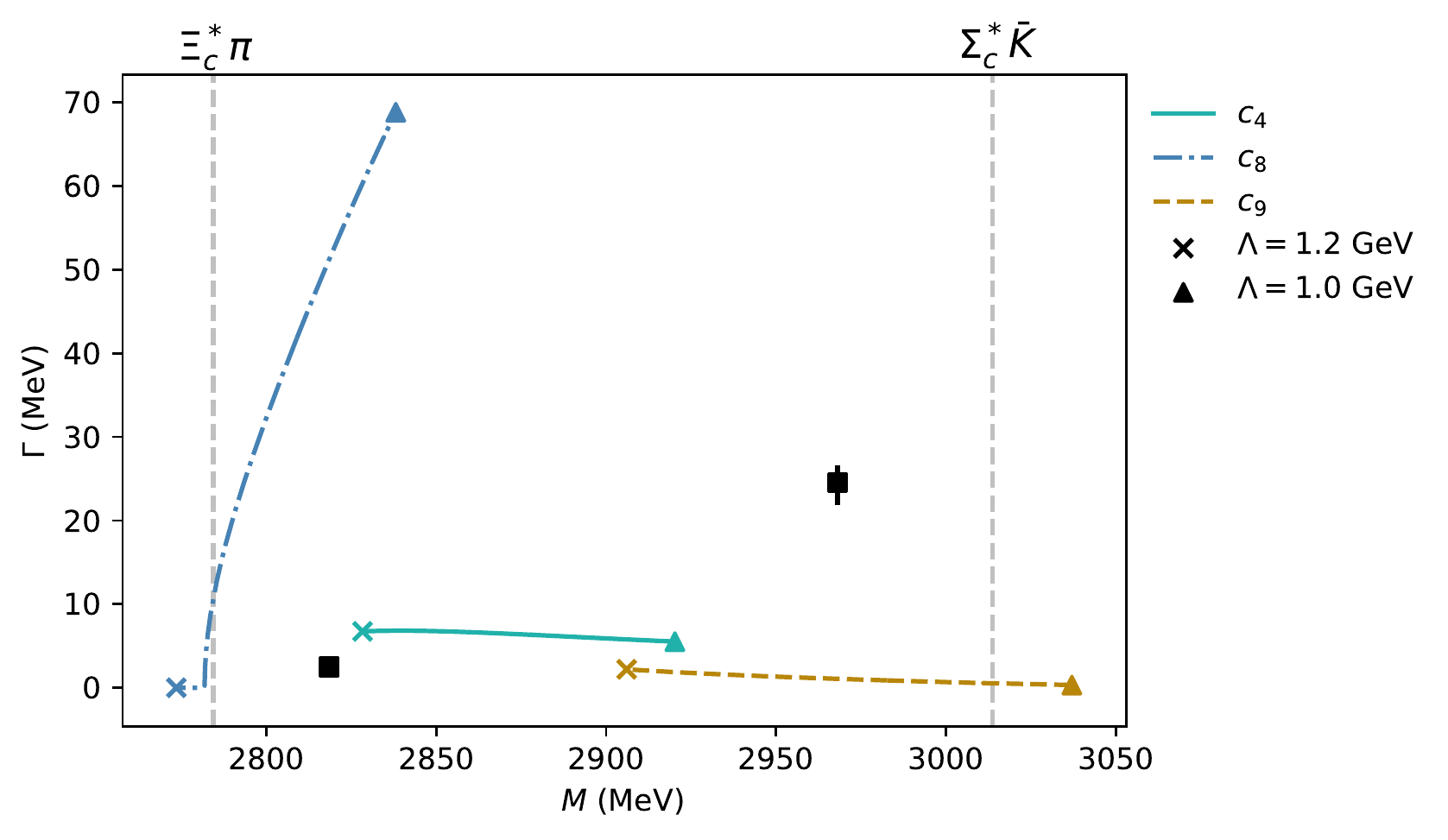}
\caption{$J=3/2$}
\label{fig:traj-c-32}
\end{subfigure}
\caption{Evolution of the masses and widths of the dynamically generated $\Xi_c$ states, as we vary the cutoff from  $\Lambda=1$ GeV (triangles) to $\Lambda=1.2$ GeV (crosses). In Fig.~\ref{fig:traj-c-12} (Fig.~\ref{fig:traj-c-32}), the  $c_1$ ($c_8$) state becomes virtual above (below) the $\Lambda_c \overline{K}$ ($\Xi_c^* \pi$) threshold. The squares and their associated errorbars show the masses and widths of the experimental $\Xi_c(2790)$ and $\Xi_c(2930)$ (Fig.~\ref{fig:traj-c-12}) 
and $\Xi_c(2815)$ and $\Xi_c(2970)$ (Fig.~\ref{fig:traj-c-32})
together with their experimental errors.  The spin-parity of both $\Xi_c(2930)$ and  $\Xi_c(2970)$ resonances are not experimentally determined~\cite{Tanabashi:2018oca}, and we have displayed them here just for illustrative purposes.}
\label{fig:traj-c}
\end{figure}

\end{widetext}

Once we have identified our $\Xi_c$ states in the cutoff scheme, we can assess the dependence of our results on this regulator, as well as their possible experimental identification. In Fig.~\ref{fig:traj-c} we show the evolution of the $c_1$ to $c_9$ states as we vary the cutoff from 1 GeV (triangles) to 1.2 GeV (crosses), and we also display different two-body  thresholds.  Moreover, in Table~\ref{tab:charmcut} we show masses and widths of the $c_1$ to $c_9$ states with $J=1/2$ or $J=3/2$, together with the couplings to the dominant baryon-meson channels ($g > 1$) and the couplings to the decay channels reported experimentally for the $\Xi_c$ states. All these results are obtained for  $\Lambda=1150$ MeV. In this table we also indicate the ${\rm SU(6)}_{\rm lsf}\times$HQSS, SU(6) and SU(3) irreducible representations (irreps), to which the $c_1$ to $c_9$ states belong (see Ref.~\cite{Romanets:2012hm} for group-structure details). 

As we evolve the cutoff value from  $\Lambda=1000$ MeV to $\Lambda=1200$ MeV, that is, from the right to left in Fig.~\ref{fig:traj-c}, we observe that some of our dynamically generated resonances can be identified with the experimental states attending to the complex energy position.

In the  $J^P=1/2^-$ sector, attending to the position in the complex plane, we observe that the $\Xi_c(2790)$ could be one of the $c_1$, $c_3$, $c_6$ or even the $c_5$ states. 
The identification with the $\Xi_c(2790)$  is possible because these states couple to $\Xi_c' \pi$, although this baryon-meson channel is not the dominant one for their dynamically generation, as seen in Table~\ref{tab:charmcut} for a $\Lambda=1150$~MeV, except for $c_5$. Indeed, this latter feature of $c_5$ disfavors its identification with  the $\Xi_c(2790)$. This is because it would become too broad ($\Gamma \ge 70$ MeV) for UV cutoffs of around 1 GeV, that would lead the $c_5$ resonance to have masses closer to the experimental one, 
as seen  in Fig.~\ref{fig:traj-c}. In addition in the DR scheme, the mass of the $c_5$ state is close to 2790 MeV, but its width is approximately of 84 MeV 
~\cite{Romanets:2012hm} (see also Fig.~\ref{fig:ctodim}), while experimentally $\Gamma_{\Xi_c(2790)}\sim 10$ MeV. 

Looking at the behavior of the $c_1$, $c_3$, $c_6$ poles with the UV cutoff  in Fig.~\ref{fig:traj-c}, it seems natural to assign the $\Xi_c(2790)$ to the $c_1$ pole. This state has a width of the order of 10 MeV for UV cutoffs in the region of 1.2 GeV, where it is located below the  $\Lambda_c\bar K$  threshold. At the same time, the state has  large $\Lambda_c\bar K$ and  small $\Xi_c'\pi$ couplings (see Table~\ref{tab:charmcut}), respectively, which explains its small experimental width despite being placed well above the latter threshold, and  it is  natural to think that the $\Lambda_c \bar K$ channel should play an important role in the dynamics of the  $\Xi_c(2790)$ given its proximity to that threshold. Note that the light degrees of freedom ({\it ldof}) in the inner structure of the $c_1$ are predominantly coupled to $j_{ldof}^\pi=0^-$ spin-parity quantum numbers~\cite{Nieves:2019nol}. Thus with this identification,   this first odd parity excited $\Xi_c$ state would not have a dominant configuration consisting of a spinless light diquark and a unit of angular momentum between it and the heavy quark, as argued for instance in the Belle Collaboration paper~\cite{Yelton:2016fqw}. This is to say, the $\Xi_c(2790)$ will not be a constituent quark model $\lambda-$mode excited state~\cite{Yoshida:2015tia,Nieves:2019nol} with $j_{ldof}^\pi=1^-$ and hence it will not form part of any HQSS doublet, thus making the assignment to $c_3$ unlikely. Actually, if the spin-parity quantum numbers for the ${\it ldof}$ in the $\Xi_c(2790)$ were predominantly $1^-$, one would expect a larger width for this resonance, since its decay to the open channel $\Xi_c' \pi$ is HQSS allowed. This is precisely the situation for the $c_3$ that is broader than the experimental state.  In summary, we conclude a large molecular $\Lambda_c \bar K$ component for the $\Xi_c(2790)$ that will have then a dominant $j_{ldof}^\pi=0^-$ configuration.  

Our present $\Xi_c(2790)$ identification with the $c_1$ pole differs from the previous assignments in Ref.~\cite{Romanets:2012hm}, where the $\Xi_c$ states were obtained using the one subtraction renormalization scheme. In this previous work,  the $c_7$ state was assigned to $\Xi_c(2790)$ due to its closeness in energy  and  the sizable $\Xi_c' \pi$ coupling within the DR scheme.

\begin{widetext}
 
\begin{table}[H]
\center
\begin{tabular}{|c|c|c|c|c||c|}
\hline
\textbf{Irreps} & \textbf{State} & $\mathbf{M}$ \textbf{(MeV)}	&	$\mathbf{\Gamma}$ \textbf{(MeV)}&	$\mathbf{J}$  & \textbf{Couplings} \\ \hline \hline
$(\mathbf{168}, \mathbf{21_{2,1}},\mathbf{3^*_2})$  &$c_1$ & $2773.59$  & $10.52$  & 1/2 & \makecell{
$g_{\Xi_c \pi} = 0.5$, 
$g_{\Xi'_c \pi} = 0.3$,
$g_{\Lambda_c	\bar{K}}	=	1.3	$,
$g_{\Sigma_c \bar{K}}=0.9$ , 
$	g_{\Lambda	D}	=	1.6	$, \\ 
$	g_{\Sigma	D}	=	1.5	$, 
$	g_{\Lambda	D^*}	=	2.9	$, 
$	g_{\Sigma	D^*}	=	1.0	$,  
$	g_{\Xi'_c	\rho}	=	1.0$, 
$g_{\Lambda_c \bar{K}^*} = 0.2$}  \\ \hline
$(\mathbf{168}, \mathbf{15_{2,1}},\mathbf{6_2})$ &$c_2$ &	$2627.5$ &  $38.84$	 & 1/2 & \makecell{$g_{\Xi_c \pi} =1.8$,
$g_{\Xi'_c \pi}=0.04$,
$g_{\Lambda_c \bar{K}} =1.2$,
$g_{\Sigma_c \bar{K}}=0.1$,
$g_{\Lambda_c \bar{K}^*} =0.04$,\\
$g_{\Sigma D} =1.2$,
$g_{\Lambda D^*} =1.0$,
$g_{\Sigma D^*} =1.9$}\\  \hline
$(\mathbf{168}, \mathbf{21_{2,1}},\mathbf{6_2})$&$c_3$ & $2790.99$ & $16.09$	 & 1/2 & \makecell{
$g_{\Xi_c \pi} = 0.3$, 
$g_{\Xi'_c \pi} = 0.8$,
$g_{\Lambda_c	\bar{K}}	=	0.2	$, 
$g_{\Sigma_c \bar{K}} =1.7$,
$g_{\Sigma D} =2.6$, \\
$g_{\Lambda D^*} =2.2$,
$g_{\Xi'_c \eta} =1.1$,
$g_{\Lambda_c \bar{K}^*} =1.0$,
$g_{\Sigma D^*} =2.3$,
$g_{\Sigma_c \bar{K}^*} =1.1$,\\
$g_{\Xi D_s^* } =1.7$}\\ \hline
$(\mathbf{168}, \mathbf{21_{2,1}},\mathbf{6_4})$&$c_4$ & $2850.89$	 & $6.76$	 & 3/2 & \makecell{
$g_{\Xi_c^* \pi}= 0.6$,
$g_{\Sigma_c^*	\bar{K}}	=	2.2$, 
$g_{\Lambda_c	\bar{K}^*}	=	1.5$, 
$g_{\Xi_c^*	\eta}	=	1.1$, 
$g_{\Sigma^*	D}	=	1.1$, \\ 
$g_{\Sigma^*	D^*}	=	1.5$, 
$g_{\Sigma_c^*	\bar{K}^*}	=	1.8$} \\ \hline
$(\mathbf{168}, \mathbf{15_{2,1}},\mathbf{3^*_2})$&$c_5$ & 	$2715.23$  & $12.28$	 & 1/2 & \makecell{
$g_{\Xi_c \pi} =0.2$,
$g_{\Xi'_c	\pi}	=	1.8$, 
$g_{\Lambda_c \bar{K}} =0.5$,
$g_{\Sigma_c	\bar{K}}	=	1.2$, 
$g_{\Lambda	D}	=	3.1$, \\
 $g_{\Lambda_c \bar{K}^*} =0.1$,
$g_{\Sigma	D}	=	1.5$} \\ \hline
$(\mathbf{120}, \mathbf{21_{2,1}},\mathbf{3^*_2})$&$c_6$ & $2807$	 &	$1.82$ & 1/2 & \makecell{
$g_{\Xi_c \pi} = 0.1$, 
$g_{\Xi'_c \pi} = 0.1$,
$g_{\Lambda_c	\bar{K}}	= 0.2$
$g_{\Sigma_c \bar{K}} =1.4$,
$g_{\Sigma D} =1.6$,\\
$g_{\Lambda D^*} =1.1$,
$g_{\Sigma D^*} =4.3$,
$g_{\Xi D_s} =1.1$,
$g_{\Sigma_c \bar{K}^*} =1.4$,
$g_{\Xi D_s^*  } =1.9$
} \\ \hline
$(\mathbf{120}, \mathbf{21_{2,1}},\mathbf{6_2})$&$c_7$ & $2922.5$	 & $2.48$	& 1/2 &  \makecell{
$g_{\Xi_c \pi} =0.2$,
$g_{\Xi'_c \pi}=0.03$,
$g_{\Lambda_c \bar{K}} =0.2$,
$g_{\Sigma_c \bar{K}}=0.1$,
$g_{\Lambda	D}	=	1.8$, \\ 
$g_{\Sigma	D}	=	1.4$, 
$g_{\Lambda	D^*}	=	1.7$, 
$g_{\Lambda_c	\bar{K}^*}	=	1.2$, 
$g_{\Sigma	D^*}	=	1.5$, 
$g_{\Xi_c	\rho}	=	1.2$,\\  
$g_{\Sigma^*	D^*	}=	3.7$, 
$g_{\Sigma_c	\bar{K}^*}	=	1.1$, 
$g_{\Xi_c^*	\rho}	=	1.0$, 
$g_{\Xi^*	D_s^*}	=	1.9$} \\ \hline
$(\mathbf{168}, \mathbf{15_{2,1}},\mathbf{3^*_4})$&$c_8$ & $2792.06$ & $22.79$	 & 3/2 & \makecell{
$g_{\Xi_c^*	\pi}	=	1.7$, 
$g_{\Sigma_c^*	\bar{K}}	=	1.0$, 
$g_{\Lambda	D^*}	=	2.4$, 
$g_{\Sigma	D^*}	=	1.2$,
$g_{\Lambda_c \bar{K}^*} = 0.2$ } \\ \hline
$(\mathbf{120}, \mathbf{21_{2,1}},\mathbf{6_4})$&$c_9$ &  $2942.05$	 &  $1.46$ 	& 3/2 & \makecell{
$g_{\Xi_c^* \pi}=0.2$,
$g_{\Sigma_c^* \bar{K}}=0.2$,
$g_{\Lambda_c \bar{K}^*} =0.4$,
$g_{\Lambda	D^*}	=	2.7$, 
$g_{\Sigma	D^*}	=	2.2$,\\  
$g_{\Sigma^*	D}	=	2.8$, 
$g_{\Sigma^*	D^*}	=	3.4$, 
$g_{\Xi^*	D_s}	=	1.4$, 
$g_{\Xi^*	D_s^*}	=	1.8$} \\ \hline
\hline
\end{tabular}
\caption{Masses and widths of the $c_1$ to $c_9$ states with $J=1/2$ or $J=3/2$ and odd parity in the $C=1$, $S=-1$ and $I=1/2$ sector, together with the couplings (in modulus) to the dominant baryon-meson channels ($g > 1$) and the couplings to the decay channels reported experimentally for the $\Xi_c$ states. All results have been 
obtained for  $\Lambda=1150$ MeV. We also indicate the ${\rm SU(6)}_{\rm lsf}\times$ HQSS, SU(6) and SU(3) irreducible representations of these states. We use the notation ${\bf R_{2J_C+1,C}}$, where ${\bf R}$ is the SU(6) irreducible representation (irrep) label (for which we use the dimension),  $J_C$ is the spin carried by the quarks with charm ($1/2$ in all cases) and $C$ the charm content (1 in all cases). In addition, we also use  ${\bf r_{2J+1}}$, where ${\bf r}$ is the SU(3) irrep, with $J$ the total angular momentum of the state (see Ref.~\cite{Romanets:2012hm} for details).}
\label{tab:charmcut}
\end{table}

\end{widetext}

With regards to the recently discovered $\Xi_c(2930)$, if we assume that this state has $J^P=1/2^-$, we could identify it either with our $c_6$ or $c_7$ states, as they both couple to the $\Lambda_c \bar K$ channel, although not dominantly as seen in Table~\ref{tab:charmcut} for a $\Lambda=1150$ MeV.  The assignment to the $c_6$ pole is, however, disfavored because of the  mass difference between this state and the experimental $\Xi_c(2930)$. As for $c_7$,  the small $\Lambda_c \bar K$ coupling of this state makes also somehow doubtful its identification with the  $\Xi_c(2930)$.  In the case of our $c_2$ and $c_5$ states, we should mention that we do not have any clear experimental candidate at this point for the $c_5$ dynamically generated $J=1/2$ state, whereas the $c_2$ state becomes broad and appears below 2650 MeV, thus not allowing for any reasonable experimental assignment.

For $J^P=3/2^-$, the analysis of the evolution of the different states in Fig.~\ref{fig:traj-c} allows for the identification of  the experimental $\Xi_c(2815)$ with $c_4$ or $c_8$. These states couple to $\Xi_c^* \pi$ in $S-$wave, although for $c_4$, couplings to other baryon-meson states ($\Sigma_c^*\bar K$, $\Lambda_c\bar K^*$ or $\Sigma_c^*\bar K^*$) are  larger as seen in Table~\ref{tab:charmcut}. The experimental $\Xi_c(2815)$ is quite narrow, $\Gamma_{\Xi_c(2815)} \sim 2-3$ MeV, despite the  $\Xi_c^*\pi$  threshold being around 30 MeV below its mass. This hints to a subdominant $\Xi_c^* \pi$ molecular component in the inner structure of this resonance. Moreover, looking at the dependence of the $J^P= 3/2^-$ pole masses and widths with the UV cutoff displayed in Fig.~\ref{fig:traj-c}, it seems reasonable to assign the $c_4$ state to the $\Xi_c(2815)$ resonance. 

As for $\Xi_c(2970)$, assuming that it has $J=3/2^-$, we could identify it with the $c_9$ state for values of the cutoff around $\Lambda \simeq 1.1$ GeV. In this case, we have to take into account that this state couples to $\Lambda_c \bar K^*$ and $\Sigma_c^* \bar K$, and $\Xi_c^* \pi$ (though not dominantly), and those baryon-meson channels can decay into $\Lambda_c \bar K \pi$ and $\Xi_c \pi \pi$, respectively. Nevertheless, the predicted width would be significantly smaller than the range of 20-30 MeV quoted in the PDG \cite{Tanabashi:2018oca} and shown in Table~\ref{tab:exp}.  Compared to the results of Ref.~\cite{Romanets:2012hm},  the $\Xi_c(2815)$ was identified  there with $c_9$, assuming that $\Xi_c(2790)$ and $\Xi_c(2815)$ were the $c_7$ and $c_9$ HQSS partners. 

In fact, we observe several HQSS partners among our states as well as possible siblings within the same $SU(3)$ representation. The $\Xi_c(2790)$ resonance belongs to an $J=1/2$ SU(3) antitriplet irrep, and it would be the ${\rm SU(6)}_{\rm lsf}\times$HQSS (see Table~\ref{tab:charmcut}) partner of a narrow $\Lambda_c^*$ state discussed in \cite{GarciaRecio:2008dp, Romanets:2012hm, Nieves:2019nol}. This latter state has large (small) $ND$ and $ND^*$ ($\Sigma_c\pi$) couplings, and  depending on the renormalization scheme (one-subtraction or UV cutoff), it is part of a double pole pattern for the $\Lambda_c(2595)$, similar to that found for the $\Lambda(1405)$ within unitarized chiral models~\cite{Oller:2000fj,GarciaRecio:2002td,Hyodo:2002pk,Jido:2003cb,GarciaRecio:2003ks,Hyodo:2011ur,Gamermann:2011mq,Kamiya:2016jqc} (see related review in \cite{Tanabashi:2018oca}), or it is located in the region of 2.8 GeV close to the $ND$ threshold~\cite{Nieves:2019nol}. 

On the other hand, the $c_3$ pole belonging to ($\mathbf{168},\mathbf{21}, \mathbf{6_2}$) representation and the $c_4$ of the ($\mathbf{168},\mathbf{21}, \mathbf{6_4}$) form a $(j_{ldof}^\pi=1^-)-$HQSS doublet. As mentioned earlier, the $c_4$ can be identified with the $\Xi_c(2815)$, but we note that the $\Xi_c(2815)$ is not the sibling of the  $\Lambda_c(2625)$ because of the different coupling strengths to $\Xi_c^* \pi$ and $\Sigma_c^* \pi$, respectively. Whereas $\Xi_c(2815)$ weakly couples to $\Xi_c^* \pi$, the $\Lambda_c(2625)$ strongly does to $\Sigma_c^* \pi$. However, this latter state is narrow because the $\Sigma_c^* \pi$ channel is closed (located around 30 MeV above the mass of the resonance). Indeed, recently it has been argued that the $\Lambda(2625)$ is probably a constituent three quark state \cite{Yoshida:2015tia,Nieves:2019nol}.

As for the $J=1/2$ $c_5$ and the $J=3/2$ $c_8$ states, those form part of a SU(6) ${\bf 15}-$plet, belonging to the ${\rm SU(6)}_{\rm lsf}\times$ HQSS  ($\mathbf{168},\mathbf{15}, \mathbf{3_2}$) and ($\mathbf{168},\mathbf{15}, \mathbf{3_4}$) irreps~\cite{Romanets:2012hm}. They form a HQSS doublet with $j_{\it dof}^\pi=1^-$ and hence have large couplings to $\Xi_c'\pi$  and $\Xi_c^*\pi$, respectively. Indeed, as a good approximation, they are dynamically generated by the charmed baryon--Goldstone boson interactions. These moderately broad states are in the SU(3)$_{2J+1}$ $\mathbf{3^*_2}$ and $\mathbf{3^*_4}$ irreps, which should be completed by one $J=1/2$ and one $J=3/2$ $\Lambda_c$ resonances stemming from the $\Sigma_c\pi $ and $\Sigma^*_c\pi$ chiral interactions~\cite{Lu:2014ina,Nieves:2019nol}, neglecting higher energy channels. The $J=3/2$ sibling is, however, not the $\Lambda_c(2625)$. As mentioned before, the $\Lambda_c(2625)$ is probably a quark model ($\lambda-$mode excitation) state~\cite{Yoshida:2015tia,Nieves:2019nol}. Another resonance with mass and width of around 2.7 GeV  and 60 MeV~\cite{Lu:2014ina,Nieves:2019nol}, that has not been discovered yet, would then be the SU(3) sibling of the $c_8$ state. 

The features of the  $J=1/2$ counterpart of $c_5$ in the $\Lambda_c$ sector are much more uncertain and depend on both the employed renormalization scheme and on the interplay between quark-model and baryon-meson degrees of 
freedom~\cite{Nieves:2019nol}. Thus, for instance neglecting the latter, it would appear around 2.6 GeV with a large width of 60-80 MeV because its sizable coupling to the $\Sigma_c\pi$ pair. Within the UV cutoff RS, this state can be easily moved below the $\Sigma_c\pi$ threshold and be identified with the narrow $\Lambda_c(2595)$~\cite{Lu:2014ina}. In the DR scheme advocated in Ref.~\cite{Romanets:2012hm}, this broad state, together with the $j^\pi_{\it ldof }=0^-$ narrow state mentioned above in the discussion of the $\Xi_c(2790)$, gives rise to a double pole structure for the  $\Lambda_c(2595)$.

Within the UV cutoff renormalization scheme examined here, the $(c_7,c_9)$ HQSS-doublet might correspond to the experimental $\Xi_c(2930)$  and $\Xi_c(2970)$ states. The $c_7$ state, that we have tentatively assigned to the $\Xi_c(2930)$, exhibits (Table~\ref{tab:charmcut}) moderate couplings to $\Xi_c\pi$ and  $\Lambda_c \bar K$, small ones to  $\Xi'_c\pi$ and $\Sigma_c\bar K$, and finally large couplings  to $\Lambda D^{(*)}$, $\Sigma D^{(*)}$ and $\Sigma^* D^{*}$. It belongs to a SU(3) sextet, where there is also a $\Omega_c$ state. The latter corresponds to the one labeled as {\bf d} in our previous study of the $\Omega_c$ odd-parity resonances~\cite{Nieves:2017jjx}, where it was tentatively assigned either to the $\Omega_c(3090)$ or the  $\Omega_c(3119)$ observed by the LHCb Collaboration in the $\Xi_c\bar K$ mode~\cite{Aaij:2017nav}. This is in fact consistent with what one might expect from its $c_7-$sibling couplings. Assuming the {\it equal spacing rule}  we could predict the possible existence of a $J=1/2^-$ $\Sigma_c$ state around 2800 MeV that will complete the sextet. The $\Sigma_c(2800)$ clearly fits into this picture since it is observed in the $\Lambda_c\pi$ channel~\cite{Tanabashi:2018oca}.

Recently there has been an analysis of the $\Xi_c$ sector within a baryon-meson molecular model based on local hidden gauge that implements the interaction between the $1/2^+$ and $3/2^+$ ground-state baryons with $0^-$ and $1^-$ mesons \cite{Yu:2018yxl}. The authors have found that five of their dynamically generated $\Xi_c$ states can be identified with the experimental $\Xi_c(2790)$, $\Xi_c(2930)$, $\Xi_c(2970)$, $\Xi_c(3055)$ and $\Xi_c(3080)$. Whereas the $\Xi_c(2790)$ would be a $1/2^-$ state, the $\Xi_c(2930)$, $\Xi_c(2970)$, $\Xi_c(3055)$ and $\Xi_c(3080)$ could be either $1/2^-$ or $3/2^-$ ones. Compared to this approach, our model identifies the experimental $\Xi_c(2790)$ and $\Xi_c(2930)$ as $1/2^-$ states, and the $\Xi_c(2815)$ and $\Xi_c(2970)$ as $3/2^-$. The different assignment is mainly due the distinct renormalization scheme used in the two approaches as well as the fact the interactions involving $D$ and $D^*$ and light vector mesons with baryons are not completely fixed by HQSS or chiral symmetries, thus allowing for different assumptions.

\vspace{0.5cm}

\subsection{ $\Xi_b$ excited states} 

With regards to the bottom sector,  the  $\Xi_b(6227)$ resonance has been recently measured by the LHCb experiment \cite{Aaij:2018yqz}, with   $\Gamma_{\Xi_b(6227)} \sim 18$ MeV. Its quantum numbers, though, remain unknown, whereas the observed decay channels are $\Lambda_b^0 K^-$ and $\Xi_b^0 \pi^-$ (see Table~\ref{tab:exp}).

We start again by revising the previous results of Ref.~\cite{GarciaRecio:2012db} with $B=-1$, $S=-1$, $I=1/2$ ($\Xi_b$ sector). Masses and widths of the dynamically generated states within our model using the DR scheme,  together with their irreps, spins  and  couplings to the dominant baryon-meson channels as well those for the experimental decay channels of $\Xi_b(6227)$  are shown in Table~\ref{tab:bottom}. We obtain nine states, which are the bottom counterparts of the $\Xi_c$ ones discussed in the previous subsection. Compared to Ref.~\cite{GarciaRecio:2012db}, we report here five more poles, since in that reference only SU(3) flavor partners of $\Lambda_b$ states were searched (members of antitriplet irreps). Also, two of them, the state at 6035 MeV with $J=1/2$ and the one at 6043 MeV with $J=3/2$ were wrongly assigned in Ref.~\cite{GarciaRecio:2012db} to the SU(6) {\bf 15} representation. Instead, their should belong to the SU(6) {\bf 21} representation, as seen in Table~\ref{tab:bottom}. Moreover, there is a state at 6073 MeV in Table  IV in Ref.~\cite{GarciaRecio:2012db} that does not appear in our present calculation. The  differences between of them are due to the difficulty in determining the number of states and their representations as we break the ${\rm SU(6)}_{\rm lsf}\times$HQSS symmetry to SU(3) in the bottom sector, as almost all  states have zero width and states with widths closer to zero are more difficult to follow in the complex energy plane. 

As in the $\Xi_c$ sector,  our $b_1$ to $b_9$ states using one-subtraction renormalization are too low in energy so as to assign any of them to the experimental  $\Xi_b(6227)$ state. Thus, we proceed as in the previous subsection and vary the renormalization scheme from one-subtraction to cutoff. In this manner, we identify our $b_1$ to $b_9$ states using one-subtraction renormalization with the ones within the cutoff scheme, and we study their evolution as we change the value of the cutoff. 

\begin{widetext}

\begin{figure}[H]
\centering
\begin{subfigure}[t]{0.7\textwidth}
\centering
\includegraphics[width=1.2\textwidth]{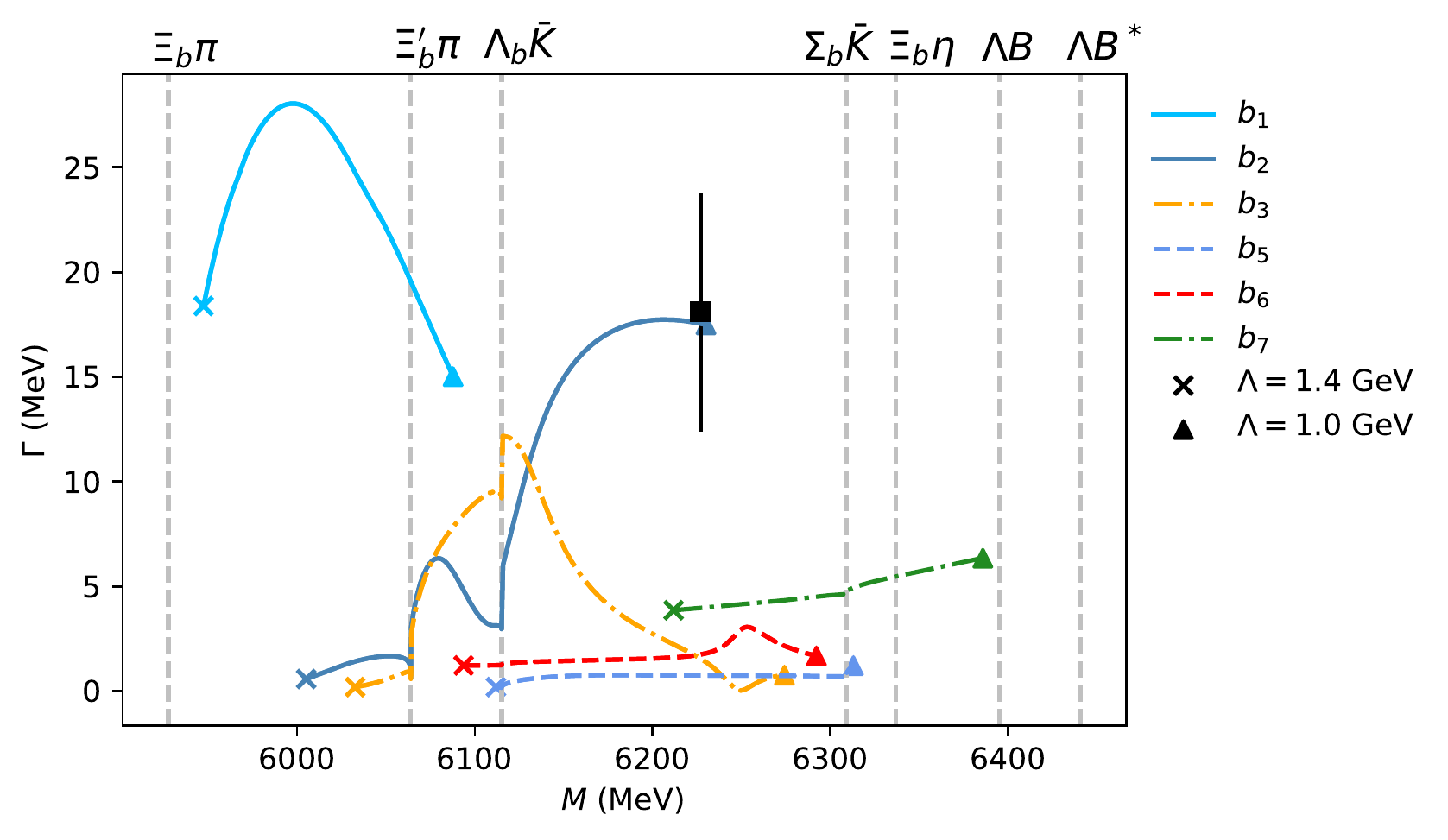}
\caption{$J=1/2$}
\label{fig:traj-b-12}
\end{subfigure}

\begin{subfigure}[t]{0.7\textwidth}
\centering
\includegraphics[width=1.2\textwidth]{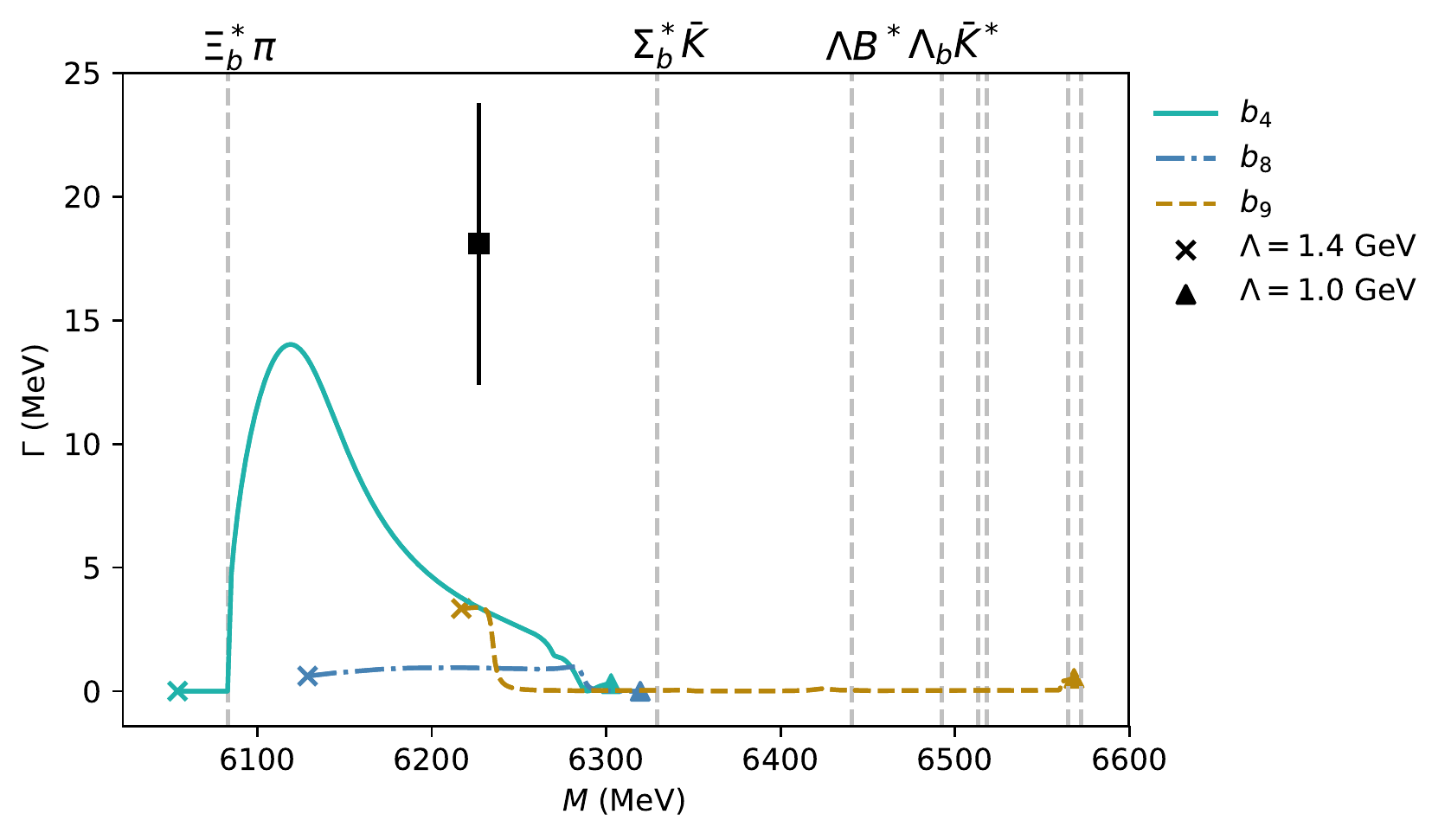}
\caption{$J=3/2$}
\label{fig:traj-b-32}
\end{subfigure}
\caption{Evolution of the masses and widths of the dynamically generated $\Xi_b$ states, as we vary the cutoff from $\Lambda=1000$ MeV (triangles) to $\Lambda=1400$ MeV (crosses), with $J=1/2$ (upper panel) and $J=3/2$ (lower panel).  The square and its bars represent the position of the $\Xi_b(6227)$ resonance, and its errors in mass and width, respectively. We show the experimental result for both values of $J$ due to its unknown quantum numbers. In Fig.~\ref{fig:traj-b-32}, the last five thresholds (not labelled in the figure because they are too close to each other) are: $\Xi_b^* \eta$ $(6492.45 \ {\rm MeV})$, $\Sigma B^*$ $(6518.35  \ {\rm MeV})$, $\Omega_b K$ $(6564.68 \  {\rm MeV})$, $\Xi_b \rho$ $(6565.04 \ {\rm MeV})$ and $\Xi_b \omega$ ($6572.12  \ {\rm MeV}$).}
\label{fig:traj-b}
\end{figure}

\end{widetext}


\begin{widetext}

\begin{table}[H]
\center
\begin{tabular}{|c|c|c|c|c||c|}
\hline
\textbf{Irreps} & \textbf{State} & $\mathbf{M_R}$ \textbf{(MeV)}	&	$\mathbf{\Gamma_R}$ \textbf{(MeV)}&	$\mathbf{J}$  & \textbf{Couplings} \\ \hline \hline
$(\mathbf{168}, \mathbf{21_{2,1}},\mathbf{3^*_2})$  &$b_1$ & $5873.98$  & $0$  & 1/2 & \makecell{$g_{\Lambda \bar{B}} = 1.3$, $g_{\Sigma \bar{B}} = 4.4$, $g_{\Lambda \bar{B}^*} = 2.3$, $g_{\Sigma \bar{B}^*} = 7.3$, $g_{\Xi \bar{B}_s} = 2.6$, \\ $g_{\Xi_b \eta '}=1.0$, $g_{\Xi \bar{B}_s^*} = 4.5$}  \\ \hline
$(\mathbf{168}, \mathbf{15_{2,1}},\mathbf{6_2})$ &$b_2$ &	$5940.85$ &  $35.59$	 & 1/2 & \makecell{$g_{\Xi_b \pi} =1.8$, $g_{\Lambda \bar{B}} =3.7$, $g_{\Lambda \bar{B}^*} =6.2$,
$g_{\Sigma \bar{B}^*} =1.6$,
$g_{\Xi \bar{B}_s} =1.1$, \\
$g_{\Xi \bar{B}_s^*   } =1.9$} \\ \hline
$(\mathbf{168}, \mathbf{21_{2,1}},\mathbf{6_2})$&$b_3$ & $5880.76$ & $0$	 & 1/2 & \makecell{$g_{\Lambda \bar{B}} =2.5$,
$g_{\Sigma \bar{B}} =2.4$,
$g_{\Lambda \bar{B}^*} =1.3$,
$g_{\Sigma \bar{B}^*} =1.6$,
$g_{\Xi \bar{B}_s} =1.7$, \\
$g_{\Sigma^* \bar{B}^*} =8.0$,
$g_{\Xi'_b \eta'} =1.0$,
$g_{\Xi^* \bar{B}_s^*} =4.9$ }\\ \hline
$(\mathbf{168}, \mathbf{21_{2,1}},\mathbf{6_4})$&$b_4$ & $5880.27$	 & $0$	 & 3/2 & \makecell{$g_{\Lambda \bar{B}^* } =2.8$,
$g_{\Sigma \bar{B}^* } =2.8$,
$g_{\Sigma^* \bar{B} } =5.0$,
$g_{\Sigma^* \bar{B}^* } =6.3$,
$g_{\Xi \bar{B}_s^* } =1.8$, \\
$g_{\Xi^* \bar{B}_s } =3.1$,
$g_{\Xi_b^* \eta' } =1.0$,
$g_{\Xi^* \bar{B}_s^* } =3.9$
} \\ \hline
$(\mathbf{168}, \mathbf{15_{2,1}},\mathbf{3^*_2})$&$b_5^*$ & 	$5949.93$  & $0.7$	 & 1/2 & $g_{\Xi'_b \pi} =1.4$,
$g_{\Lambda \bar{B}} =6.2$,
$g_{\Lambda \bar{B}^*} =3.8$,
$g_{\Sigma^* \bar{B}^*} =1.6$,
$g_{\Xi^* \bar{B}_s^*} =2.2$ \\ \hline
$(\mathbf{120}, \mathbf{21_{2,1}},\mathbf{3^*_2})$&$b_6$ & $6034.80$	 &	$28.8$ & 1/2 & \makecell{$g_{\Xi_b \pi} =1.0$,
$g_{\Lambda_b \bar{K}} =2.0$,
$g_{\Lambda \bar{B}} =1.0$,
$g_{\Lambda \bar{B}^*} =2.1$,
$g_{\Sigma \bar{B}^*} =1.1$, \\
$g_{\Xi \bar{B}_s} =1.3$,
$g_{\Xi \bar{B}_s^*   } =2.1$} \\ \hline
$(\mathbf{120}, \mathbf{21_{2,1}},\mathbf{6_2})$&$b_7$ & $6035.39$	 & $0.02$	& 1/2 &  \makecell{$g_{\Sigma_b  \bar{K}} =2.3$,
$g_{\Lambda \bar{B}} =1.0$,
$g_{\Sigma \bar{B}} =4.5$,
$g_{\Sigma \bar{B}^*} =2.8$,
$g_{\Xi_b \omega} =1.2$, \\
$g_{\Sigma^* \bar{B}^*} =2.3$} \\ \hline
$(\mathbf{168}, \mathbf{15_{2,1}},\mathbf{3^*_4})$&$b_8^*$ & $5958.20$ & $0$	 & 3/2 & \textbf{-- R.S. not connected --} \\ \hline
$(\mathbf{120}, \mathbf{21_{2,1}},\mathbf{6_4})$&$b_9$ &  $6043.28$	 &  $0$ 	& 3/2 & \makecell{$g_{\Sigma_b^* \bar{K} } =2.3$,
$g_{\Lambda \bar{B}^* } =1.1$,
$g_{\Sigma \bar{B}^* } =5.5$,
$g_{\Sigma^* \bar{B} } =1.5$,
$g_{\Xi_b \omega } =1.2$, \\
$g_{\Sigma^* \bar{B}^* } =1.7$} \\ \hline

\end{tabular}
\caption{Masses and  widths of the $b_1$ to $b_9$ states with $J=1/2$ or $J=3/2$ in the $B=-1$, $S=-1$ and $I=1/2$ sector, together with the couplings to the dominant baryon-meson channels and the couplings to the experimental decay channels of the $\Xi_b(6227)$,  using one-subtraction renormalization, as in Table IV of Ref.~\cite{GarciaRecio:2012db}. We also indicate the ${\rm SU(6)}_{\rm lsf}\times$ HQSS, SU(6) and SU(3) irreducible representations of these states, as in Table~\ref{tab:charmcut}. States with $*$ are virtual states. Note that the $b_8^*$ lies in the real axis, but in a sheet that is not connected to the physical sheet, thus we are not showing the couplings indicating "R.S (real sheet) not connected".}
\label{tab:bottom}
\end{table}
\end{widetext}

In  Fig.~\ref{fig:traj-b} we display the evolution of the masses and widths of the dynamically generated $\Xi_b$ states as we vary the cutoff from $\Lambda=1000$ MeV (triangles) to  $\Lambda=1400$ MeV, for $J=1/2$ (upper plot)  and $J=3/2$ (lower plot).  The square and its bar represent the position of the $\Xi_b(6227)$ resonance, and the error for its mass and width, respectively. We show the experimental result ($\Xi_b(6227)$) for both $J=1/2$ and $J=3/2$ because its quantum numbers have not been determined yet. Additionally,  in Table~\ref{tab:bottomcut}, we collect  the masses and the widths of the $b_1$ to $b_9$ states with $J=1/2$ or $J=3/2$, together with the couplings to the dominant baryon-meson channels and the couplings to the decay channels of the $\Xi_b(6227)$,  for $\Lambda=1150$ MeV as in the charm sector. We also indicate the ${\rm SU(6)}_{\rm lsf}\times$ HQSS, SU(6) and SU(3) irreducible representations of these states. 

We might try now to assign the experimental $\Xi_b(6227)$ to any of our states, while determining the negative parity baryons with $B=-1$ belonging to the same ${\bf 3^*}$ and ${\bf 6}$ SU(3) representations. The observed decay modes,  $\Lambda_b^0 K^-$, $\Xi_b^0 \pi^-$~\cite{Aaij:2018yqz}, of the resonance support that this state should have $1/2^-$ spin-parity, assuming $S-$wave. Moreover, the $j^\pi_{\it ldof}=0^-$ component should be also quite relevant, which according to the couplings collected in Table~\ref{tab:bottomcut} makes plausible its identification either with the $b_1$ or $b_2$ states. The evolution displayed in the upper plot of Fig.~\ref{fig:traj-b} leads us to assign the $\Xi_b(6227)$ to the $b_2$ state. 
The $b_2$ pole  would stem from a SU(6) {\bf 15-}plet, composed of $J=1/2$ and $J=3/2$ SU(3) antitriplets  and of a $J=1/2$ SU(3) sextet, where the $\Xi_b(6227)$ would be accommodated. The $J=1/2^-$ $\Lambda_b(5912)$ and  $J=3/2^-$  $\Lambda_b(5920)$ (LHCb \cite{Aaij:2012da}) would be part of the ${\bf 3^*_2}$ and ${\bf 3^*_4}$ multiplets forming a HQSS-doublet~\cite{GarciaRecio:2012db}. These antritriplets should be completed by another HQSS-doublet of $\Xi_b$ and $\Xi^*_b$ states, $b_5$ and $b_8$, that according  to Fig.~\ref{fig:traj-b} and Table~\ref{tab:bottomcut} should have masses around 6250 MeV and could be seen in the $\Sigma_b^{(*)}\bar K$ and $\Xi_b^{(\prime *)}\pi$ modes.

Coming back to the $\Xi_b(6227)$,  it belongs to a $j_{ldof}^\pi=0^--$sextet that should be completed by $J=1/2$ $\Sigma_b$ and  $\Omega_b$  states. The recent  $\Sigma_b(6097)$
resonance seen by the LHCb Collaboration~\cite{Aaij:2018tnn}  in the $\Lambda_b\pi$ channel nicely fits in this multiplet. Relying again in the {\it equal spacing rule}, we could foresee the existence of  a $J=1/2$ $\Omega_b$ odd parity state  with a mass of around 6360 MeV that should be observed in the $\Xi_b\bar K$ channel. 
Some molecular $\Omega_b$ states were predicted previously in Ref.~\cite{Liang:2017ejq}, but all of them above 6.4 GeV.

Previous works based on molecular approaches have also found the $\Xi_b(6227)$ as a dynamically-generated state. In Refs.~\cite{Lu:2014ina,Huang:2018bed} a unitarized model using the leading-order chiral Lagrangian found the $\Xi_b(6227)$ as a $S$-wave $\Sigma_b \bar K$ molecule, with a preferred $1/2^-$ spin-parity assignment \cite{Huang:2018bed}. In our present model the $\Lambda \bar B^*$, $\Sigma \bar B$ and  $\Lambda \bar B$ are the dominant channels in the generation of the $\Xi_b(6227)$, though it also couples (weakly) to $\Sigma_b \bar K$. The main difference between models comes from the fact that our scheme has a more extensive number of channels, whereas the antitriplet and sextet multiplets of ground-state baryons mix when constructing the interaction matrices. Also, the work of Ref.~\cite{Yu:2018yxl} has also analyzed the $\Xi_b$ sector. The authors have found two poles with masses close to the $\Xi_b(6227)$ and widths $\sim 25-30$ MeV, close to the experimental one, with $1/2^-$ and $3/2^-$ spin-parity. In our model we identify the $\Xi_b(6227)$ as a $1/2^-$ state and, again, the difference arises because of the renormalization scheme and the interaction matrices involving $D$, $D^*$ and light vector mesons.

\begin{widetext}

\begin{table}[H]
\center
\begin{tabular}{|c|c|c|c|c||c|}
\hline
\textbf{Irreps} & \textbf{State} & $\mathbf{M_R}$ \textbf{(MeV)}	&	$\mathbf{\Gamma_R}$ \textbf{(MeV)}&	$\mathbf{J}$  & \textbf{Couplings } \\ \hline \hline
$(\mathbf{168}, \mathbf{21_{2,1}},\mathbf{3^*_2})$  &$b_1$ & $6025.46$  & $25.88$  & 1/2 & \makecell{
$g_{\Xi_b \pi} = 0.94$,
$g_{\Lambda_b \bar{K}} =1.4$,
$g_{\Xi_b \eta} =2.1$,
$g_{\Sigma \bar{B}} =1.4$,
$g_{\Sigma \bar{B}^*} =2.6$,\\
$g_{\Xi \bar{B}_s^*   } =1.3$}  \\ \hline
$(\mathbf{168}, \mathbf{15_{2,1}},\mathbf{6_2})$ &$b_2$ &	$6152.61$ &  $15.29$	 & 1/2 & \makecell{
$g_{\Xi_b \pi} = 0.33$,
$g_{\Lambda_b \bar{K}} = 0.51$,
$g_{\Sigma_b \bar{K}} = 0.40$,
$g_{\Lambda \bar{B}} =1.9$,
$g_{\Sigma \bar{B}} =2.1$,\\
$g_{\Lambda \bar{B}^*} =7.3$,
$g_{\Xi \bar{B}_s} =1.6$} \\ \hline
$(\mathbf{168}, \mathbf{21_{2,1}},\mathbf{6_2})$&$b_3$ & $6179.4$ & $3.81$	 & 1/2 & \makecell{
$g_{\Xi_b \pi} = 0.05$,
$g_{\Lambda_b \bar{K}} = 0.1$,
$g_{\Lambda \bar{B}} =1.08$,
$g_{\Sigma \bar{B}} =1.92$,
$g_{\Lambda \bar{B}^*} =1.87$,\\
$g_{\Omega_b K} =2.26$,
$g_{\Xi \bar{B}_s} =5.13$, 
$g_{\Xi \bar{B}_s^*   } =2.65$,
$g_{\Xi_b \phi  } =2.29$,
$g_{\Omega_b K^*} =1.04$,\\
$g_{\Xi'_b \phi} =1.15$}\\ \hline
$(\mathbf{168}, \mathbf{21_{2,1}},\mathbf{6_4})$&$b_4$ & $6202.73$	 & $4.48$	 & 3/2 & \makecell{ 
$g_{\Lambda \bar{B}^* } =2.3$,
$g_{\Sigma \bar{B}^* } =1.5$,
$g_{\Omega_b K } =2.2$,
$g_{\Xi \bar{B}_s^* } =5.5$,
$g_{\Xi_b \phi } =2.3$, \\
$g_{\Omega_b \bar{K}^* } =1.2$,
$g_{\Xi_b^* \phi       } =1.3$} \\ \hline
$(\mathbf{168}, \mathbf{15_{2,1}},\mathbf{3^*_2})$&$b_5$ & 	$6243.02$  & $0.74$	 & 1/2 & \makecell{
$g_{\Xi_b \pi} = 0.02$,
$g_{\Lambda_b \bar{K}} =0.12$, 
$g_{\Sigma_b \bar{K}} =0.48$, 
$g_{\Sigma \bar{B}} =1.8$,
$g_{\Sigma \bar{B}^*} =6.9$} \\ \hline
$(\mathbf{120}, \mathbf{21_{2,1}},\mathbf{3^*_2})$&$b_6$ & $6212.26$	 &	$1.6$ & 1/2 & \makecell{
$g_{\Xi_b \pi} = 0.05$,
$g_{\Lambda_b \bar{K}} = 0.01$,
$g_{\Sigma_b \bar{K}} =1.2$,
$g_{\Lambda \bar{B}} =1.3$,
$g_{\Sigma \bar{B}} =4.9$,\\
$g_{\Lambda \bar{B}^*} =2.3$,
$g_{\Xi'_b \eta} =1.6$} \\ \hline
$(\mathbf{120}, \mathbf{21_{2,1}},\mathbf{6_2})$&$b_7$ & $6327.28$	 & $5.29$	& 1/2 &  \makecell{
$g_{\Xi_b \pi} = 0.01$,
$g_{\Lambda_b \bar{K}} =0.02$, 
$g_{\Lambda \bar{B}} =1.4$,
$g_{\Sigma \bar{B}} =1.3$,
$g_{\Lambda \bar{B}^*} =1.2$,\\
$g_{\Lambda_b \bar{K}^*} =1.9$,
$g_{\Sigma \bar{B}^*} =1.3$, 
$g_{\Xi_b \rho} =1.5$,
$g_{\Sigma^* \bar{B}^*} =2.2$} \\ \hline
$(\mathbf{168}, \mathbf{15_{2,1}},\mathbf{3^*_4})$&$b_8$ & $6240.82$ & $0.92$	 & 3/2 & \makecell{$g_{\Xi_b^* \pi } =0.15$, $g_{\Sigma_b^* \bar{K} } =1.3$,
$g_{\Lambda \bar{B}^* } =2.0$,
$g_{\Xi_b^* \eta } =1.5$,
$g_{\Sigma \bar{B}^* } =4.8$} \\ \hline
$(\mathbf{120}, \mathbf{21_{2,1}},\mathbf{6_4})$&$b_9$ &  $6459.42$	 &  $0.02$ 	& 3/2 & \makecell{$g_{\Xi^* \bar{B}_s } =4.5$,
$g_{\Omega_b K^* } =2.2$,
$g_{\Xi'_b \phi } =3.0$,
$g_{\Xi^* \bar{B}_s^* } =3.0$,
$g_{\Omega_b \bar{K}^* } =1.0$, \\
$g_{\Xi_b^* \phi       } =1.3$} \\ \hline
\end{tabular}
\caption{As Table~\ref{tab:charmcut}, but for the $\Xi_b$ sector ($\Lambda=1150$ MeV). }
\label{tab:bottomcut}
\end{table}
\end{widetext}


\begin{widetext}

\begin{figure}[H]
\begin{subfigure}{0.5\textwidth}
\centering
\scalebox{1.}{
\begin{tikzpicture}
\begin{feynman}
\node[draw,dot,fill] (a0);
\vertex[below = 1cm of a0] (v);
\node[draw,dot,fill,left = .8cm of v] (a1);
\node[draw,dot,fill,right = .8cm  of v] (b1);
\vertex[right = 1 cm of b1] (b2) {$\Xi_b(6240)$ ?};
\vertex[right = 1.5 cm of a0] (v2) {$\Lambda_b(5912)$};

\diagram* {
(a1),
(b1),
(v),
};
\end{feynman}
\end{tikzpicture}
}\\
\end{subfigure}%
\begin{subfigure}{0.5\textwidth}
\centering
\scalebox{1.}{
\begin{tikzpicture}
\begin{feynman}
\node[draw,dot,fill] (a0);
\vertex[below = 1cm of a0] (v);
\node[draw,dot,fill,left = .8cm of v] (a1);
\node[draw,dot,fill,right = .8cm  of v] (b1);
\vertex[right = 1 cm of b1] (b2) {$\Xi_b(6240)$ ?};
\vertex[right = 1 cm of a0] (v2) {$\Lambda_b(5920)$};

\diagram* {
(a1),
(b1),
(v),
};
\end{feynman}
\end{tikzpicture}
}
\end{subfigure}
\caption{Bottom baryon states classified within the $J=1/2$ (left diagram) and $J=3/2$ (right diagram) SU(3) ${\bf 3^*}$ irreps. The question mark indicates states predicted in this work.}
\label{fig:fdres32y34}
%
\vspace{1cm}
\begin{subfigure}{0.5\textwidth}
\centering
\scalebox{1.}{
\begin{tikzpicture}
\begin{feynman}
\node[draw,dot,fill] (u0);
\node[draw,dot,fill,left=1cm of u0] (u2);
\node[draw,dot,fill,right=1cm of u0] (u1);
\vertex[below = 1cm of u0] (v);
\node[draw,dot,fill,left = .5cm of v] (a1);
\node[draw,dot,fill,right = .5cm  of v] (b1);
\vertex[right = 1.2 cm of b1] (b2) {$\Xi_c(2930)$};
\vertex[right = 1.2 cm of u1] (v2) {$\Sigma_c(2800) $};
\node[draw,dot,fill,below = 1 cm of v] (d0);
\vertex[right = 2 cm of d0] (v3) {$\Omega_c(3090)/\Omega_c(3119)$};

\diagram* {
(a1),
(b1),
(v),
(d0),
};
\end{feynman}
\end{tikzpicture}
}\\
\end{subfigure}%
\begin{subfigure}{0.5\textwidth}
\centering
\scalebox{1.}{
\begin{tikzpicture}
\begin{feynman}
\node[draw,dot,fill] (u0);
\node[draw,dot,fill,left=1cm of u0] (u2);
\node[draw,dot,fill,right=1cm of u0] (u1);
\vertex[below = 1cm of u0] (v);
\node[draw,dot,fill,left = .5cm of v] (a1);
\node[draw,dot,fill,right = .5cm  of v] (b1);
\vertex[right = 1.2 cm of b1] (b2) {$\Xi_b(6227) $};
\vertex[right = 1.2 cm of u1] (v2) {$\Sigma_b(6097) $};
\node[draw,dot,fill,below = 1 cm of v] (d0);
\vertex[right = 1.2 cm of d0] (v3) {$\Omega_b(6360)$?};

\diagram* {
(a1),
(b1),
(v),
(d0),
};
\end{feynman}
\end{tikzpicture}
}
\end{subfigure}
\caption{Charm and bottom resonances classified within SU(3) ${\bf 6}$ irreps with $J=1/2$, which however stem from different ${\rm SU(6)}_{\rm lsf}\times$HQSS irreps: $(\mathbf{120}, \mathbf{21},\mathbf{6_2})$ and  $(\mathbf{168}, \mathbf{15},\mathbf{6_2})$, respectively. The question mark indicates states predicted in this work.}
\label{fig:fdres62}
\end{figure}
\end{widetext}

\section{Conclusions}
 \label{conc}

In this work we have explored the possible molecular interpretation of several experimental excited $\Xi_c$ and $\Xi_b$ states. We have used a coupled-channel unitarized model, that is based on a ${\rm SU(6)}_{\rm lsf}\times$HQSS-extended WT baryon-meson interaction, within the on--shell approximation. We have paid a special attention to the dependence of our predictions on the renormalization scheme, so as to assess the robustness of our results.

We have presented a molecular interpretation for the experimental $\Xi_c(2790)$, $\Xi_c(2815)$, $\Xi_c(2930)$, $\Xi_c(2970)$ and $\Xi_b(6227)$ states, and have predicted the spin-parity quantum numbers of the latter three resonances. We have found that the $\Xi_c(2790)$ state has a large molecular $\Lambda_c \bar K$ component, with a dominant $j_{\it ldof}^\pi=0^-$ configuration, and  discussed the differences between the $3/2^-$ $\Lambda_c(2625)$ and $\Xi_c(2815)$ states, finding that they cannot be SU(3) siblings.  We have also predicted the existence of other $\Xi_c-$states, not experimentally detected yet, being two of them siblings of the two poles that form the $\Lambda_c(2595)$. Interestingly, the recently discovered $\Xi_c(2930)$ and $\Xi_c(2970)$ are found to be HQSS partners.

The flavor-symmetry content of the framework has also  allowed us to understand the nature of the $\Sigma_c(2800)$ and $\Sigma_b(6097)$ states, for which we have determined their spin-parity.  Moreover, we have predicted several states, some of them displayed in Figs.~\ref{fig:fdres32y34} and ~\ref{fig:fdres62} (marked with a ? symbol).  Among them, we stress the $\Omega_b(6360)$ state, with a dominant $\Xi_b \bar K$ contribution, in the sextet where the $\Sigma_b(6097)$ and $\Xi_b(6227)$ are located, together with the $\Xi_b(6240)$ and $\Xi_b^*(6240)$ states,  partners of the HQSS doublet $\Lambda_b(5912)$ and $\Lambda_b(5920)$ discussed in \cite{GarciaRecio:2012db}.

\section{Acknowledgements}
L.T. acknowledges support from Deutsche Forschungsgemeinschaft under
Project Nr. 383452331 (Heisenberg Programme) and Project Nr. 411563442.
R. P. Pavao wishes to thank the Generalitat Valenciana in the program GRISOLIAP/2016/071. This
research is supported by the Spanish Ministerio de Econom\'ia y
Competitividad and the European Regional Development Fund, under
contracts FIS2017-84038-C2-1-P, FPA2013-43425-P,
FPA2016-81114-P and SEV-2014-0398;  the THOR COST Action CA15213; and by the EU STRONG-2020 project under the program
H2020-INFRAIA-2018-1, grant agreement no. 824093.

\bibliography{ref}

\end{document}